\documentclass[twocolumn,showpacs,amsmath,amssymb,floatfix]{revtex4}
\usepackage{graphicx}
\usepackage{dcolumn}
\usepackage{bm}

\begin{document}

\title{Percolation transition in networks with degree-degree correlation}

\author{Jae Dong Noh}
\affiliation{Department of Physics, University of Seoul,
  Seoul 130-743, Korea}

\date{\today}
\begin{abstract}
We introduce an exponential random graph model for networks 
with a fixed degree distribution and with a tunable degree-degree correlation. 
We then investigate the nature of a percolation transition in
the correlated network with the Poisson degree distribution. 
It is found that 
negative correlation is irrelevant in that the percolation transition in the
disassortative network belongs to the same universality class of the
uncorrelated network. Positive correlation turns out to be
relevant. The percolation transition in the assortative network is
characterized by the non-diverging mean size of finite clusters and 
power-law scalings of the density of the largest cluster and the cluster
size distribution in the non-percolating phase as well as at the critical
point. Our results suggest that the unusual type percolation transition in
the growing network models reported recently may be inherited from the
assortative degree-degree correlation. 
\end{abstract}
\pacs{89.75.Hc, 64.60.-i, 05.70.Fh, 05.50.+q}
\maketitle

\section{Introduction}\label{sec1}
Percolation in complex networks have been attracting a lot of 
interest in the statistical physics 
community~\cite{Stauffer&Aharony94,Albert02}. A network may undergo a phase
transition as one discards nodes or links successively. 
When the fraction of remaining nodes or links is greater than a threshold, 
the network possesses a giant cluster which consists of a finite fraction of
interconnected nodes. 
In the opposite case the giant cluster disappears and all nodes disintegrate 
into small clusters. It is called the percolation phase transition that
separates the two phases.
The percolation transition in complex networks, as well as in regular
lattices~\cite{Stauffer&Aharony94}, 
is interesting because of its relevance to robustness of network systems
against random failure and epidemic
spreading~\cite{Albert00,Callaway00,Newman01,Cohen00,Cohen_etal02,Lee04,Lee05,Derenyi05,Goltsev06,Serrano06}.

The random network of Erd\H{o}s and R\'{e}nyi~(ER) is a prototypical model
for complex networks~(see Ref.~\cite{Albert02} for review). 
An ER network with $N$ nodes is constructed by linking 
each pair of nodes with the probability $p/[(N-1)/2]$, 
or by adding $pN$ links between randomly selected pairs of nodes. 
The link density is given by $p$, and the degree distribution follows the
Poisson distribution $P_{deg.}(k) =  e^{- \langle k\rangle} \langle
k\rangle^k / k!$ with the mean degree $\langle k\rangle = 2p$.

The ER network is {\it uncorrelated} in the sense that it lacks any 
structural correlation. This property allows one to study the
percolation transition analytically. We summarize some known results:
(i)~The percolation order parameter $P$ is defined as the probability 
that a node 
belongs to a giant cluster.
It exhibits a threshold behavior with the
power-law scaling 
\begin{equation}\label{eq:Pbeta}
P \sim (p - p_c)^\beta
\end{equation}
for $p>p_c=1/2$ with the exponent $\beta = 1$. 
(ii)~Let $n(s)$ be the number of clusters of size $s$ per node. 
At the critical point it
follows the power-law distribution
\begin{equation}\label{eq:ns}
n(s) \sim s^{-\tau}
\end{equation}
with the exponent $\tau = 5 / 2$.  
For $p\neq p_c$ it does not follow the power law.
(iii)~The mean cluster size $S$ is defined as the average
size of {\it finite} clusters reached from nodes selected randomly.
It also displays the power-law scaling
\begin{equation}\label{eq:Sgamma}
S \sim | p-p_c |^{-\gamma}
\end{equation}
with the exponent $\gamma=1$. Note that the percolation transition belongs
to the same universality class as the mean field
transition~\cite{Stauffer&Aharony94}.

The study has been extended to
scale-free networks with the power-law degree distribution $P_{deg.}(k) \sim
k^{-\lambda}$. Making use of the generating function
method~\cite{Callaway00,Newman01,Cohen_etal02} or the mapping to the
$q= 0$ limit of the $q$-state Potts model~\cite{Lee04,Lee05}, 
researchers find that the percolation transition in uncorrelated
scale-free networks is characterized by the
power-law scalings with the $\lambda$-dependent exponents.

Recently the percolation transition in a class of growing networks draws
interest~\cite{Callaway01,Dorogovtsev01,JKim02,Krapivsky04}. 
The common feature of these networks is that the numbers of nodes and links
are increasing in time with the density of links $p$ fixed. 
Adding a node and making a link corresponds to nucleation of a cluster and
merging of clusters, respectively. As one varies $p$, finite clusters 
condense into a giant cluster giving rise to the percolation 
transition~\cite{BenNaim07}.  Interestingly the nature
of the transition is different from that observed in the
uncorrelated networks. The critical properties are summarized in the
following: (i)~The percolation order parameter exhibits an
essential singularity 
\begin{equation}\label{Pessential}
P \sim \exp\left[ - \frac{a}{\sqrt{p-p_c}} \right] 
\end{equation}
with a constant $a$.
(ii)~The cluster size distribution follows the power law $n(s)\sim
s^{-\tau}$ in the whole phase 
with $p\leq p_c$. The exponent value is varying with $p$ and takes 
$\tau=3$ at the critical point with a logarithmic correction. 
(iii)~The mean size of finite clusters
$S$ does not diverge at the critical point. Instead, it is finite and 
shows a discontinuous jump at $p=p_c$. 
These features are reminiscent of the
Berezinskii-Kosterlitz-Thouless~(BKT) transition in two dimensional 
equilibrium systems with continuous symmetry~\cite{BKT}. However, 
there is no similarity in the underlying mechanism for the transitions.

Previous studies reveal that there exist at least two distinct universality 
classes for the percolation transition in complex networks.  One is
characterized with the power-law singularity and the other 
with the essential singularity. It raises a question for the 
key ingredient that is responsible for the universality class. Similarly one
may ask a question whether the essential singularity is observed in a
non-growing network. 

There is an important observation that the growing
networks~\cite{Callaway01,Dorogovtsev01,JKim02,Krapivsky04} have a positive
degree-degree correlation. A positive degree-degree correlation, or
phrased as an assortative mixing,
refers to the tendency toward making links between nodes of 
similar degrees~\cite{Newman02}. Consider a pair of connected nodes in a 
growing network. As a network grows, the two nodes acquire more and more links 
generating a positive correlation. On the other hand, those networks
displaying the power-law type percolation transition do
not have any degree correlation. It gives a hint that the degree 
correlation may be an important factor determining the universality class.
In this work, we will investigate the effect of the degree correlation 
on the nature of the percolation transition. 

The degree correlation of a network can be quantified by the 
assortativity~\cite{Newman02}
\begin{equation}\label{assortativity}
r = \frac{ \langle kk'\rangle_{l} - \langle(k+k')/2\rangle_{l}^2}
    { \langle( k^2 + k'^2)/2\rangle_{l} - \langle (k+k')/2\rangle_{l}^2} \ ,
\end{equation}
where $\langle \cdot\rangle_{l}$ denotes the average over all links and 
$(k,k')$ denotes the degrees of two nodes at either end of links. 
Its sign indicates a positive~(assortative) or negative~(disassortative)
degree correlation. It vanishes for uncorrelated~(neutral) networks.
The degree correlation can also be monitored with the 
nearest neighbor degree $K_{nn}(k)$ which is given by the average degree of
neighbors of degree-$k$ nodes~\cite{Pastor-Satorras01,Vazquez02}. 
It is an increasing~(decreasing) function of
$k$ for networks with a positive~(negative) correlation.
Several characteristics of complex networks with the degree correlation have 
been studied~\cite{Vazquez03,Brunet03,Brunet04,Bianconi06}. 
However, the universality class of the percolation transition 
has not been understood yet. 

In this work, we will investigate the sole
effect of the degree correlation on the nature of the percolation transition.
It necessitates a model for networks with a tunable degree correlation to a
given fixed degree distribution. In order to avoid an interference with any
other ingredient, the model is required to be random in other aspects than
the degree distribution and the degree correlation. We propose such a model
in Sec.~\ref{sec2}. It belongs to a class of the exponential random
graph model~\cite{Park04}. In this class, a network model is defined as
a Gibbsian ensemble of networks with an associated network Hamiltonian.
The model and its structural properties 
will also be studied. In Sec.~\ref{sec3}, we will investigate the percolation
transition of the model as varying the degree correlation.
Summary and discussions will be given in Sec.~\ref{sec4}.

\section{Exponential random graph model}\label{sec2}
The statistical ensemble approach is useful in modeling a network with a 
specific property~\cite{Berg02,Palla04,Park04,Biely06}.
Suppose that one wants to construct a network model which is specified 
by an observable $x$. It is suggested that~\cite{Park04} 
such a model can be defined as the Gibbsian ensemble over the set of 
networks ${\cal G}=\{G\}$ with the probability distribution
\begin{equation}\label{Gibbsian}
P(G) \propto {e^{-H(G)}} \ .
\end{equation}
Here $H(G)$, called the network Hamiltonian, is given by
\begin{equation}
H(G) = \theta x(G) \ .
\end{equation}
The value of the observable $x$ can be adjusted by the parameter 
$\theta$ through the relation
\begin{equation}
x = \sum_{G\in {\cal G}} P(G) x(G) \ .
\end{equation}
This is called the exponential random graph~(ERG) model.

Our purpose is to construct an ERG model for networks with a 
fixed degree distribution $P_{deg.}(k)$ and 
with a tunable degree correlations.  Then it may be natural 
to use the assortativity $r$ in Eq.~(\ref{assortativity}) 
for the network Hamiltonian $H$.  
One can find a simpler form by using the requirement that the
degree distribution $P_{deg.}(k)$ should be fixed. 
Note that $\langle (k+k')/2\rangle_l$ and $\langle (k^2+k'^2)/2\rangle_l$ 
are constants to a given degree distribution.
Hence it suffices to consider the term $\langle kk'\rangle_l$ only in
Eq.~(\ref{assortativity}) for the Hamiltonian.

Following is the formal definition of our model: Let ${\cal G}'$ be
the set of $N$-node networks that are specified by a degree 
distribution $P_{deg.}(k)$. 
A network $G$ is conveniently described with the 
adjacency matrix $\mathbf{A}=(a_{ij})~(i,j=1,\cdots,N)$ whose matrix element 
takes the value $a_{ij}=1 \mbox{\ or\ } 0$ if nodes $i$ and $j$ are 
connected or not. 
The model is defined as the Gibbsian ensemble over
${\cal G}'$ with the network Hamiltonian given by
\begin{equation}\label{JHamiltonian}
H(G) = -\frac{J}{2} \sum_{i,j=1}^N a_{ij} k_i k_j \ ,
\end{equation}
where $k_i = \sum_j a_{ij}$ denotes the degree of a node $i$ and
$J$ is a control parameter. A positive~(negative) correlation is
favored by a positive~(negative) value of $J$. 
The model may have any degree distribution. However, we only consider the
simplest Poisson distribution as the ER network since we are interested in
the effect of the degree correlation.

\begin{figure}[t]
\includegraphics[width=0.8\columnwidth]{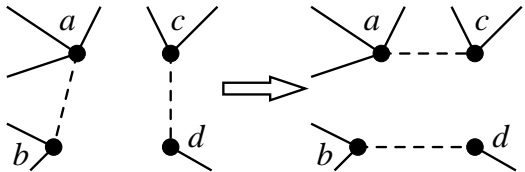}
\caption{This picture shows a link rewiring process. Two links between
node pairs $(a,b)$ and $(c,d)$ are chosen at random. They are then
rewired to connect pairs $(a,c)$ and $(b,d)$.}
\label{fig1}
\end{figure}
The Gibbsian ensemble can be simulated by using a Monte Carlo method.
We start with an ER network with $N$ nodes and $L=p_0N$ links, and update
network configurations via the so-called link rewiring
process~\cite{Maslov02} as illustrated in Fig.~\ref{fig1}.
A link rewiring trial from a configuration $G$ to $G'$ is accepted
with the probability $\min\left\{ 1,e^{-(H(G')-H(G))}\right\}$. 
Then, the Monte Carlo dynamics 
leads to the Gibbsian ensemble in the stationary state.
It is noteworthy that the link rewiring process preserves the degree of
each node.
Therefore the ERG model combined with the Monte Carlo method allows us to
study properties of complex networks with a given degree distribution but
with a different degree correlation. The degree correlation can be adjusted
with the parameter $J$.

\begin{figure}[t]
\includegraphics*[width=\columnwidth]{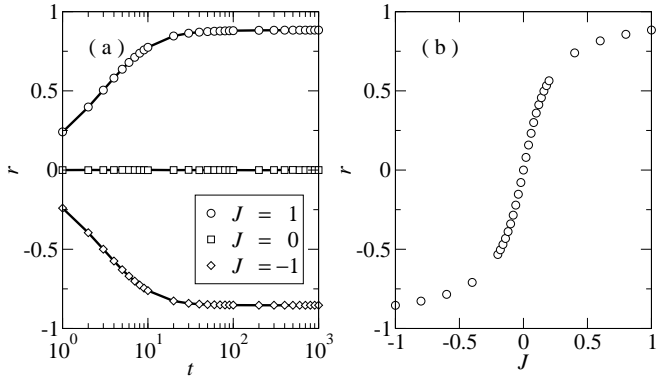}
\caption{(a)~Time evolution of the assortativity in networks with 
$N=16000$~(solid lines) and $32000$~(symbols) nodes. (b)~Stationary state
values of the assortativity as a function of $J$ in networks with $N=32000$
nodes.}
\label{fig2}
\end{figure}
Our model has a finite relaxation time. We tested relaxation dynamics
at $N=16000$ and $32000$ and $p_0=2$. Figure~\ref{fig2}(a) shows the time
evolution of the assortativity $r$, averaged over $N_S=100$ samples,
at $J=1$, $0$, and $-1$. One finds that the assortativity converges to
stationary state values in finite Monte Carlo steps with a negligible finite
size effect. 

The stationary state value of the assortativity at $p_0=2$ 
is presented in Fig.~\ref{fig2}(b), which was measured with 
$N=32000$. We find that the assortativity vanishes at $J=0$ and is 
positive for $J>0$ and negative for $J<0$. 
At $J=0$, links are rewired randomly, which is supposed to lead to an
uncorrelated network~\cite{Maslov02}. The assortativity measure confirms
the expectation. A positive~(negative) value of $J$ leads to an
assortative~(disassortative) network.

\begin{figure}[t]
\includegraphics[width=\columnwidth]{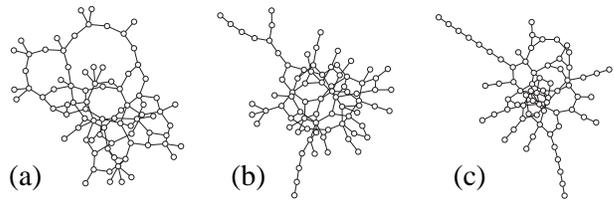}
\caption{Snapshots of networks with $J=-1$ in (a), $J=0$ in (b), and $J=1$
in (c). $N=100$ and $p_0=1$.}
\label{fig3}
\end{figure}
Typical network configurations are shown in Fig.~\ref{fig3}.
They are obtained from the Monte Carlo simulations with $J=-1$, 0, and 1,
respectively, starting with the same initial ER network with $N=100$ and 
$p_0=1$. Shown are only the largest cluster in each case. In the
disassortative case~($J=-1$), most of large degree nodes with $k>2$ 
are paired with small degree nodes with $k=1$. On the contrary, in the
assortative case~($J=1$), large degree nodes and small degree nodes 
are segregated from each other. While large degree nodes form an 
interwoven core, small degree nodes form branches emanating from the core.
The neutral network~($J=0$) shows the features of the assortative and the
disassortative networks simultaneously. We note that the assortative network
has the most inhomogeneous structure for the segregation.

\begin{figure}[t]
\includegraphics[width=\columnwidth]{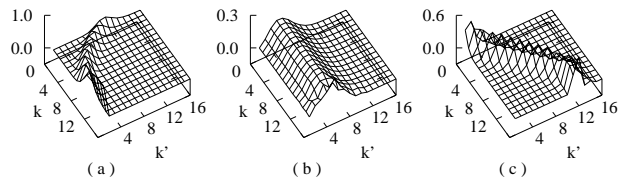}
\caption{The conditional probability distribution $P(k'|k)$ for networks
with $N=10^4$ and $p_0 = 2.0$. $J=-1$ in (a), $J=0$ in (b), and $J=1$
in (c).}
\label{fig4}
\end{figure}
The degree correlation can also be seen from the probability
distribution $p(k'|k)$~\cite{Newman02}. 
It is the conditional probability that a node at one end of a randomly chosen
link has degree $k'$ provided that a node at the other end has degree $k$.
We find that it, being viewed as a function of $k'$, is sharply peaked 
for all values of $J$. The peak position $k'_{peak}$ decreases, remains at a
constant value, or increases when $J<0$, $J=0$, or $J>0$, respectively.
Numerical data showing these behaviors are presented in Fig.~\ref{fig4}.


\section{Percolation transition}\label{sec3}
We proceed to study the percolation transition in our network model in the
following manner:
(i)~An ER network is prepared with an initial link density $p_0=2$.
(ii)~A correlated network out of the ER network is generated 
by applying the Monte Carlo dynamics to a given value of $J$. 
(iii)~Links are selected at random and removed successively. 
In the mean while, percolation properties such as
the density of the largest cluster $P$, the average size of finite 
clusters $S$, and the cluster size distribution $n(s)$ are measured as
functions of remaining link density $p$.
Those procedures are repeated $N_S = {\cal O}(10^3)$ times, 
and all measurements are averaged over those samples. 

We remark on the effect of the random link removal on degree correlation.
Consider an arbitrary network with a link density $p_0$ and an assortativity
$r_0$. Assume that the link density becomes $p$ after the random link
removal. A straightforward algebra shows that the assortativity $r$ of the
link-removed network is given by
\begin{equation}\label{r_tr}
r = \frac{r_0}{ 1 + \frac{1-p/p_0}{p/p_0} 
\left( \frac{\langle{k^2}\rangle/\langle{k}\rangle-1}
            {\langle{k^3}\rangle/{\langle{k}\rangle}- 
              (\langle{k^2}\rangle/\langle{k}\rangle)^2}
\right)}
\end{equation}
where $\langle k^n\rangle$ is the $n$-th moment of the degree of the
initial network~\cite{unpub}. 

Our networks before the random link removal have the
the Poisson degree distribution
$P_{deg.}(k) = e^{-\langle k \rangle} \langle k \rangle^k / k!$ 
with the mean degree $\langle k\rangle = 2 p_0$. So the
moments are given by $\langle k^2\rangle = \langle k \rangle^2 + \langle k
\rangle$ and $\langle k^3
\rangle = \langle k \rangle^3 + 3\langle k \rangle^2 + \langle k \rangle$.  
Inserting these into Eq.~(\ref{r_tr}), one obtains that $r = (p/p_0) r_0$. 
This relation guarantees that a network remains
to be assortative~(disassortative) during random link removals 
if it is assortative~(disassortative) initially.

\begin{figure}[t]
\includegraphics*[width=\columnwidth]{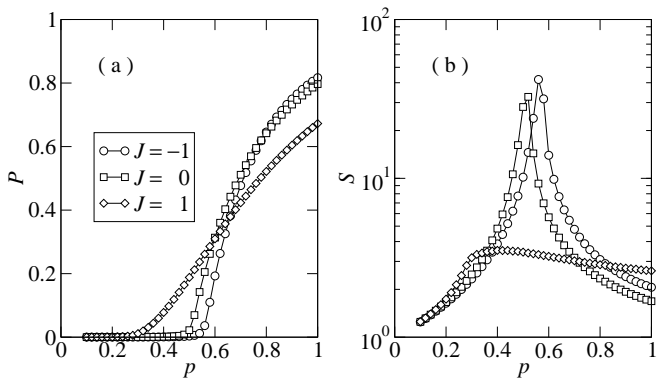}
\caption{The density of the largest cluster $P$ in (a) and the mean size of
finite clusters $S$ in (b). These are obtained for the networks with
$N=8\times 10^4$ nodes averaged over $N_S = 1000$ samples.}
\label{fig5}
\end{figure}
We have studied numerically the percolation transition in the networks at 
several values of $J$. It seems that the nature of the percolation
transition depends only on the sign of $J$. So, we will present the results 
for the cases with $J=-1$, $0$, and $1$ as the representatives of 
disassortative, neutral, and assortative networks, respectively. 

In Fig.~\ref{fig5}, we compare the density of the largest cluster $P$
and the mean size of finite clusters $S$. For all values of $J$, 
the quantity $P$ displays a threshold behavior indicating the percolation 
transition at a nonzero value of $p$. There are noticeable differences. 
The giant cluster shows up earliest
in the assortative network~($J=1$). 
It, however, grows so slowly that it becomes
smaller than those in the neutral~($J=0$) and the disassortative~($J=-1$)
network at large $p$. These properties can be understood from the typical
configurations given in Fig.~\ref{fig3}. An assortative network
consists of a highly interconnected core and branches emanating from it.
The core is stable against the random link removal, whereas the branches 
can be easily disconnected from the core. 
Apart from the quantitative features, the data for $P$ also suggest that
the scaling behavior of $P$ near the percolation threshold may be dependent
on the assortativity.

The behavior of $S$ shows even more conspicuous difference.
There are sharp peaks near the percolation threshold for $J=0$ and $J=-1$.
However, the assortative network with $J=1$ does not exhibit such a peak.
This is reminiscent of the percolation transition in the growing
networks~\cite{Callaway01,Dorogovtsev01,JKim02,Krapivsky04}.

Numerical data in Fig.~\ref{fig5}(a) and (b) indicate that the degree
correlation may affect the nature of the percolation transition. We will
investigate the nature of the percolation transition in each case using a
finite size scaling~(FSS) method.

For finite values of $N$, the scaling law in Eq.~(\ref{eq:Pbeta}) for $P$
has the FSS form
\begin{equation}\label{Pfss}
P(p,N) = N^{-\beta/\bar{\nu}}\ {\cal P}\left((p-p_c)N^{1/\bar{\nu}}\right) 
\ ,
\end{equation}
where $\bar{\nu}$ is the FSS exponent. The scaling
function ${\cal P}(x)$ has the limiting behavior ${\cal P}(x\gg 1)
\sim x^{\beta}$ and ${\cal P}(x \rightarrow 0 )=c_1$ with a constant $c_1$.
Similarly the scaling law in Eq.~(\ref{eq:Sgamma}) for $S$ has
the FSS form
\begin{equation}\label{Sfss}
S(p,N) = N^{\gamma/\bar{\nu}} \ {\cal S}\left( (p-p_c)
N^{1/\bar{\nu}}\right) \ .
\end{equation}
The scaling function ${\cal S}(x)$ has the limiting behavior ${\cal
G}(|x| \gg 1) \sim x^{-\gamma}$ and ${\cal S}(|x| \rightarrow 0)=c_2$ with a
constant $c_2$.

At $J=0$, our model is equivalent to the ER random network. 
It is known that the percolation threshold 
is located at $p_c=1/2$. The critical exponents are given by those of 
the mean field theory, that is, $\beta = \beta_{MF} = 1$ and 
$\gamma = \gamma_{MF} = 1$~\cite{Stauffer&Aharony94,Albert02}. 
There is a little subtlety in the FSS
exponent $\bar{\nu}$. It was conjectured that $\bar{\nu}$ is given by the
product of the mean field correlation length exponent $\nu_{MF}$ and the
upper critical dimensionality $d_u$ provided that the criticality belongs to
the mean field universality class~\cite{hpark07}. The conjecture yields that
$\bar{\nu}=3$, which is indeed the case for the ER random network~\cite{Lee04}.
\begin{figure}[t]
\includegraphics*[width=\columnwidth]{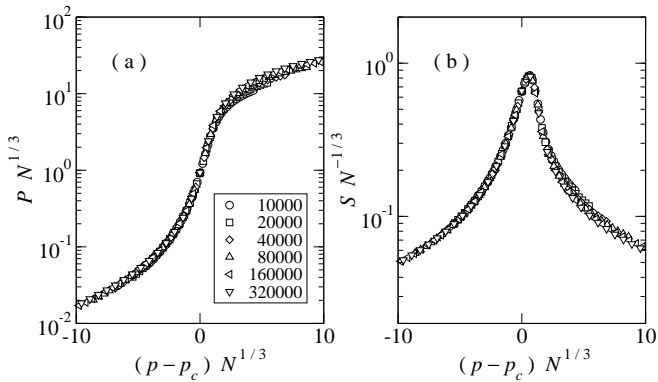}
\caption{Scaling plots for $P$ and $S$ at $J=0$.}
\label{fig6}
\end{figure}
In order to test the FSS ansatz, we have performed a scaling analysis.
Figure~\ref{fig6} shows the scaling plots for $P$ and $S$ according to the
FSS forms in Eqs.~(\ref{Pfss}) and (\ref{Sfss}) with $p_c = 1/2$ 
and the mean field exponents $\beta = 1$, $\gamma=1$, and $\bar{\nu}=3$.
All data taken from different network sizes $N$ collapse onto single curves 
quite well indicating the validity of the FSS form and the critical exponents.

\subsection{Disassortative networks ($J=-1$)}\label{sec3.1}
\begin{figure}[t]
\includegraphics*[width=\columnwidth]{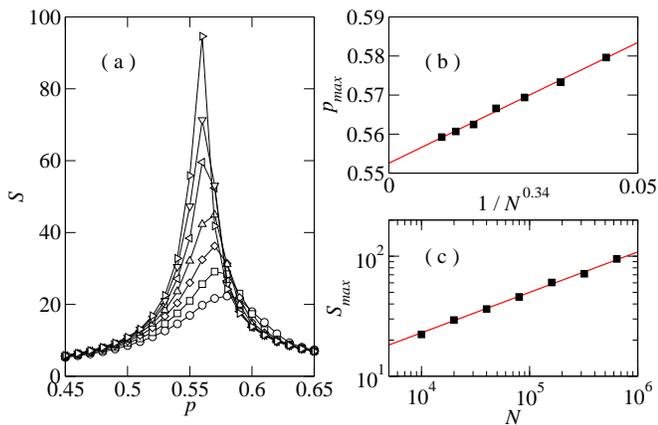}
\caption{Numerical results at $J=-1$. (a) The mean size of finite clusters
at various networks sizes $N$ ranging from $N=10^4$~(circles) to 
$N=64\times10^5$~(right-triangles). (b) The peak positions $p_{peak}$ vs
$1/N^{0.340}$. The equation for the straight line is $y=0.553+0.628x$.
(c) The peak height $S_{max}$ vs. $N$. The straight line has the slope
$0.340$.} 
\label{fig7}
\end{figure}

In this subsection, we will investigate the nature of the percolation
transition in the disassortative network with $J=-1$. 
In order to locate the percolation threshold $p_c$, we focus on the FSS
behavior of $S$ plotted in Fig.~\ref{fig7}(a).
It is evident that there are peaks which become sharp as
$N$ increases. If the percolation transition is characterized by the 
power-law type singularity, the FSS form in Eq.~(\ref{Sfss}) implies that 
the peak position $p_{max}$ approaches the critical point $p_c$ as
\begin{equation}\label{pc_fss}
p_{max} = p_c + a N^{-1/{\bar{\nu}}}
\end{equation}
with a constant $a$ and that the peak height $S_{max}$ grows as
\begin{equation}\label{Smax}
S_{max} \sim N^{\gamma/\bar{\nu}} \ .
\end{equation}

\begin{figure}[t]
\includegraphics*[width=\columnwidth]{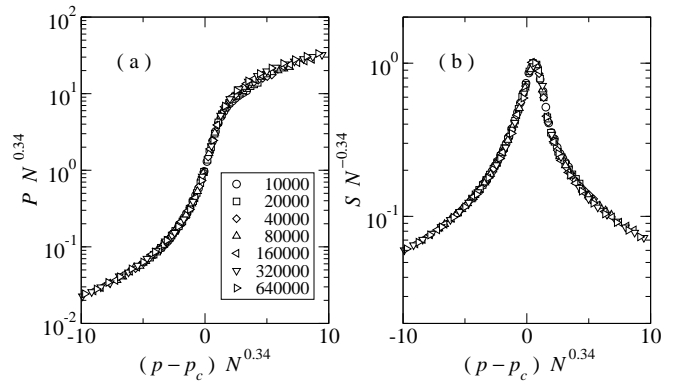}
\caption{Scaling plots for $P$ in (a) and $S$ in (b) at $J=-1$.}
\label{fig8}
\end{figure}

We fitted the data $S$ near the peak to a quadratic function to interpolate
the values of $p_{max}$ and $S_{max}$ at each value of $N$. 
Thus obtained values of $p_{max}$ and $S_{max}$ are fitted well to
Eqs.~(\ref{pc_fss}) and (\ref{Smax})~[see Figs.~7(b) and (c)], from which
we find that 
\begin{equation}
p_c = 0.553(5),\ \ 1/\bar{\nu} = 0.34(1),\ \ \gamma/\bar{\nu} = 0.34(1).
\end{equation}
According to the FSS form in Eq.~(\ref{Pfss}), one expects that 
the largest cluster $P$ scales algebraically as
\begin{equation}
P \sim N^{-\beta/\bar{\nu}}
\end{equation}
at the critical point $p=p_c$. Fitting data near the critical point,
we obtained that
\begin{equation}
\beta/ \bar{\nu} = 0.34(1).
\end{equation}
Figure~\ref{fig8} shows that all data for $P$ and $S$ 
at different values of $N$ collapse on to single curves, which proves the
reliability of the numerical results for the critical exponents.

The critical behaviors and the values of the critical exponents 
are compatible with those of random networks. 
Therefore we conclude that the
percolation transition in the disassortative network belongs to the same
universality class as that in the uncorrelated neutral network.

\subsection{Assortative networks ($J=1$)}\label{sec3.2}
\begin{figure}[t]
\includegraphics*[width=\columnwidth]{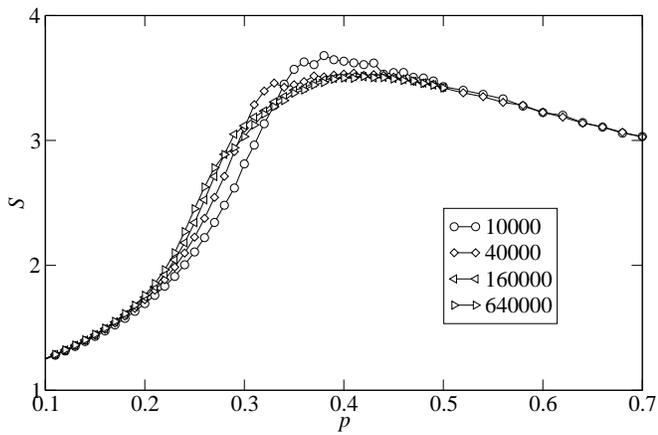}
\caption{The mean sizes of finite clusters in the assortative networks with
$J=1$ at different values of $N$.}
\label{fig9}
\end{figure}

In this subsection we will investigate the nature of the percolation 
transition in the assortative network.
We have already noticed from Fig.~\ref{fig5} that the assortative network
behaves differently from the neutral and the disassortative network.  
The difference is stressed again in Fig.~\ref{fig9}, where we present
numerical data for $S$
obtained from networks at different sizes $N=10^4$,$\cdots$,$64\times 10^4$.
Although there is a peak, it does not sharpen as $N$ increases. At the same
time, finite size effects are non-negligible near the peak.
This may be regarded as an indication that the assortative network 
does not undergo a percolation transition at all. It may be 
another possible scenario that there is a percolation transition associated
with non-divergent $S$.

\begin{figure}[t]
\includegraphics*[width=\columnwidth]{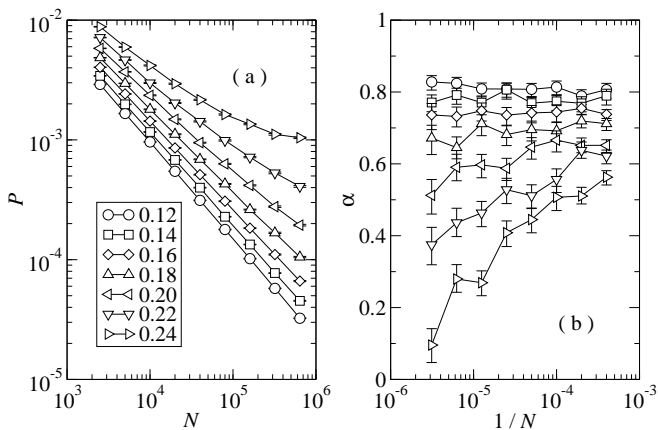}
\caption{(a)~$P$ versus $N$ at several values of
$p=0.12$,$\cdots$,$0.24$. (b)~Effective exponent $\alpha$ versus $1/N$.
}
\label{fig10}
\end{figure}

We study FSS behaviors of the percolation order parameter $P$.
The FSS behaviors plotted in Fig.~\ref{fig10} clearly shows that the network
undergoes the percolation transition at finite $p_c$. 
As $N$ increases, $P(N)$ approaches a constant value for large $p$ 
while it follows a power-law decay $P(N) \sim N^{-\alpha}$ 
for small $p$.  
We make use of an effective exponent $\alpha$ defined as
$$\alpha(N) = -\frac{\ln\left[ P(2N)/P(N)\right]}{\ln 2}$$
in order to locate the transition point. From the effective exponent plot
in Fig.~\ref{fig10}(b), we estimate that the transition point is at $p_c =
0.20(2)$. At the critical point, the density of the largest cluster 
follows the power-law scaling $P\sim N^{-\alpha_c}$ with 
\begin{equation}
\alpha_c = 0.6(1) \ .
\end{equation}

Note that the exponent $\alpha_c$ is distinct from the corresponding value
$(\beta/\bar{\nu})=1/3$ for the uncorrelated neutral network. Note also that
$S$ does not diverge at the percolation threshold. 
Based on these evidences, we conclude that the percolation transition in
the assortative network belongs to a distinct universality class.
The assortativity is an essential ingredient for the universality class of
the percolation transition.

\begin{figure}[t]
\includegraphics*[width=\columnwidth]{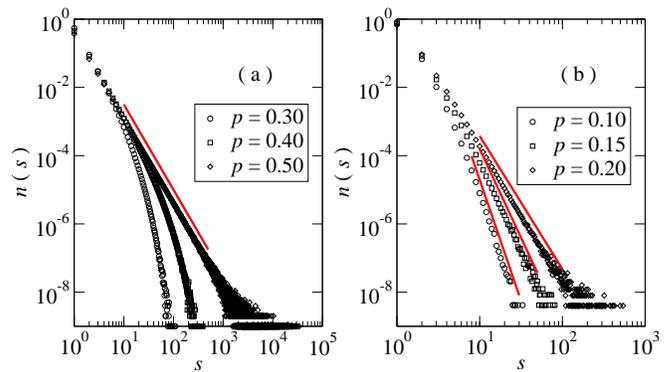}
\caption{The cluster size distributions in the neutral networks~($J=0$) in
(a) and the assortative networks~($J=1$) in (b) of size $N=10^6$. 
The straight line in (a) has the slope 5/2, 
and the straight lines in (b) have the slope 3.9, 5.2, 7.1, respectively.}
\label{fig11}
\end{figure}
Our remaining task is to characterize the percolation transition in the
assortative network. Figure~\ref{fig10} shows that $P$ follows the
power-law scaling $P\sim N^{-\alpha}$ not only at the critical point at
$p=p_c$ but also in the non-percolating phase at $p<p_c$. 
Furthermore the value of the exponent $\alpha$ is varying with $p$.
It implies that the system is in a critical state for $p\leq p_c$. The
criticality is also observed in the power-law scaling of
the cluster size distribution $n(s)\sim s^{-\tau}$ with the continuously 
varying exponent $\tau$ in the non-percolating phase. Figure~\ref{fig11}
compares the cluster size distribution in the neutral network and the
assortative network. In the neutral network, the cluster size distribution
follows the power law only at the critical point with $\tau = 5/2$. On the
other hand, it follows the power law both at and below the critical point in
the assortative network. At the critical point, the exponent is given by
$\tau\simeq 3.9$.

The power-law scaling behaviors of $P$ and $n(s)$ implies that the system is
critical in the non-percolating phase. Hence the percolation transition cannot
be described by power-law type scaling laws. 
Instead, the assortative network model shares many features in common with the
growing network models~\cite{Callaway01,Dorogovtsev01,JKim02,Krapivsky04} 
in regard to the critical behaviors. 
The non-divergence of $S$ at the critical point and
the power-law scaling of $P\sim N^{-\alpha}$ and $n(s)\sim s^{-\tau}$ in the
non-percolating phase are such common features. 
At the critical point our numerical estimates are $\alpha\simeq 0.6$ 
and $\tau\simeq3.9$, while the corresponding values are 
$\alpha=1/2$ and $\tau = 3$ in the growing network model~\cite{JKim02}.
We attribute these discrepancies to the logarithmic corrections at the
critical point~\cite{JKim02}.
Our model is a generic one for networks with assortative degree correlation.
Therefore our numerical results
suggest that assortative degree correlation is responsible for 
the unusual scaling behaviors observed in the growing network models.

\section{Summary and discussion}\label{sec4}
In summary, we have investigated numerically the nature of the 
percolation transition in networks with degree correlation. As a model 
for the correlated networks, we have introduced the exponential random 
graph model
with the Hamiltonian given in Eq.~(\ref{JHamiltonian}) under the restriction
that the degree distribution is fixed. 
Using the model combined with the Monte Carlo method explained in
Sec.~\ref{sec2}, one can generate correlated networks to a given degree 
distribution~(taken as the Poisson distribution in this work).
Numerical results show that the negative degree correlation is irrelevant in
that the disassortative network exhibits the same type percolation transition
as the neutral network. On the other hand the positive correlation turns
out to be relevant. The percolation transition in the assortative network is
characterized by the non-diverging $S$ at $p=p_c$ and power-law scaling of
$P \sim N^{-\alpha}$ and $n(s)\sim s^{-\tau}$ with the continuously
varying exponents $\alpha$ and $\tau$ in the non-percolating phase.

The scaling behaviors of the assortative network
are compatible with those of the growing network
models~\cite{Callaway01,Dorogovtsev01,JKim02,Krapivsky04}. It strongly
suggests that the unusual percolation transition in the growing network
models is inherited from the assortative degree correlation.
This conclusion is highly plausible but not decisive yet.
It is worthwhile to mention a discrepancy in the property of the mean size
of finite clusters $S$. The growing network models show a discontinuous 
jump in $S$ at $p=p_c$. However we do not find an indication of
such a discontinuity
in the assortative network. It remains as an unsolved question whether
the discontinuous jump in $S$ is an universal property or not.
Numerical studies were limited to networks up to size $N=64\times 10^4$
because the Monte Carlo dynamic generating correlated networks is slow.
Numerical data up to that size fail to justify exclusively
the essential singularity in $P$ as in Eq.~(\ref{Pessential}).
In this respect, it is desirable to find an efficient algorithm 
with which one can generate the correlated network of much larger sizes.
At the same time, it will help us understand better the property of 
correlated networks if one can find an analytically tractable model.
These are left for future studies.

Acknowledgement: This work was supported by Korea Research Foundation Grant
funded by the Korean Government (MOEHRD, Basic Research Promotion Fund)
(KRF-2006-003-C00122). The author thanks H. Park and B. Kahng for helpful
discussions.

\end{document}